\documentclass[12pt]{article}

\begin{document}
\pagenumbering{arabic}
\begin{titlepage}
\title{Conformally invariant teleparallel theories of gravity}

\author{J. W. Maluf$\,^{(a)}$ and  F. F. Faria$\,^{(b)}$ \\
Instituto de F\'{\i}sica, \\ 
Universidade de Bras\'{\i}lia, \\
C.P. 04385 \\
70.904-970 Bras\'{\i}lia, DF, Brazil\\}

\date{}
\maketitle
\begin{abstract}
We analyze the construction of conformal theories of gravity in the realm of
teleparallel theories. We first present a family of conformal theories which
are quadratic in the torsion tensor and are constructed out of the tetrad 
field and of a scalar field. For a particular value of a coupling constant, 
and in the gauge where the scalar field is restricted to assume a constant
value, the theory reduces to the teleparallel equivalent of general
relativity, and the tetrad field satisfies Einstein's equations. A second 
family of theories is formulated out of the tetrad field only, and the theories
are not equivalent to the usual Weyl Lagrangian. Therefore the latter is not 
the unique genuinely geometrical construction that yields a conformally 
invariant action. The teleparallel framework allows more possibilities for
conformal theories of gravity.
\end{abstract}
\vfill
\par
\bigskip
\noindent (a) wadih@unb.br, jwmaluf@gmail.com\par
\noindent (b) felfrafar@hotmail.com\par
\end{titlepage}
\bigskip
\section{Introduction}
Conformal invariance is a natural extension of the symmetries of Einstein's
general relativity, namely, the covariance of the metric tensor and of 
Einstein's equations under space-time coordinate transformations. The 
invariance of a possible theory of gravity under conformal transformations 
requires either that one adds a scalar field to the theory, or adopt the 
Lagrangian based on the square of Weyl's tensor. The addition of a scalar 
field to the Hilbert-Einstein action was already considered by Dirac 
\cite{Dirac} and Utyiama \cite{Utyiama}. They established conformal theories
of gravity in terms of the metric tensor, and of a vector and scalar fields.
Dirac \cite{Dirac} noted that the addition of a scalar field to the Lagrangian
density would avoid the complications that arise with Weyl's action principle,
that yields field equations much more complicated than Einstein's equations. 

A conformal theory of gravity is a viable candidate for the quantum theory
of gravity \cite{tHooft1,tHooft2}, since it is expected to be
renormalizable. In the latter references a conformal transformation is 
performed in the non-conformally invariant Hilbert-Einstein action such that 
the conformal factor $\omega(x)$ is extracted from the metric tensor 
$g_{\mu\nu}$ according to $g_{\mu\nu}=\omega^2\,\hat{g}_{\mu\nu}$,
and considered as an independent degree of freedom. In ref. \cite{tHooft1} it
is suggested the condition $\det\hat{g}=-1$ on $\hat{g}_{\mu\nu}$.
Because of this condition, the quantity $\hat{g}_{\mu\nu}$ is not exactly a 
tensor, since it does not transform as an ordinary tensor. But it is argued 
that the effective theory constructed out of $\hat{g}_{\mu\nu}$ is conformally
(scale) invariant.

A theory of gravity with conformal symmetry has been addressed recently in
connection with the dark matter and dark energy problems 
\cite{Mannheim1,Mannheim2}, and in the investigation of anisotropic
cosmological solutions \cite{Moon2}.
In this approach the Lagrangian density is given
by the square of Weyl's tensor, and is ultimately written in terms of the
square of the Ricci tensor and of the scalar curvature. It is usually argued
in the literature that Weyl's Lagrangian density is the unique combination 
constructed out of the curvature tensor that leads to a conformally invariant
theory. While this statement might be correct, Weyl's Lagrangian is not the
unique genuinely geometrical quantity that displays conformal invariance.

In this article, we show that it is possible to establish a conformally (scale)
invariant family of theories in the framework of teleparallel theories of 
gravity constructed out of the tetrad field. We will show that a quadratic 
combination of the torsion tensor that slightly deviates from the Lagrangian
density of the teleparallel equivalent of general relativity (TEGR) can be 
made conformally invariant, with the help of a scalar field. The addition of
a suitable kinetic term for the scalar field yields a theory that reduces 
to the TEGR when the scalar field is gauge fixed to a constant value. We
remark that the role of conformal transformations in $f(T)$ type theories has 
already been analyzed in ref. \cite{Yang}.

We will also show that a family of Lagrangian 
densities that depend on the fourth
power of the torsion tensor, with no addition of a scalar field, is invariant 
under conformal transformations and yields conformal theories of gravity that
so far have not been presented and investigated in the literature. These 
theories are not equivalent to the usual Weyl's theory. Since the theories 
presented here are invariant under conformal transformations, a possibility 
exists that they could contribute to the formulation of the quantum theory of 
gravity. 

\bigskip
Notation: space-time indices $\mu, \nu, ...$ and SO(3,1) indices $a, b, ...$
run from 0 to 3. Time and space indices are indicated according to
$\mu=0,i,\;\;a=(0),(i)$. The tetrad field is denoted $e^a\,_\mu$, and the 
torsion tensor reads $T_{a\mu\nu}=\partial_\mu e_{a\nu}-\partial_\nu e_{a\mu}$.
The flat, Minkowski spacetime metric tensor raises and lowers tetrad indices 
and is fixed by $\eta_{ab}=e_{a\mu} e_{b\nu}g^{\mu\nu}= (-1,+1,+1,+1)$. The 
determinant of the tetrad field is represented by $e=\det(e^a\,_\mu)$.

The torsion tensor defined above is often related to the object of
anholonomity $\Omega^\lambda\,_{\mu\nu}$ via 
$\Omega^\lambda\,_{\mu\nu}= e_a\,^\lambda T^a\,_{\mu\nu}$. However, we assume
that the (non-scale invariant) spacetime geometry is defined by the tetrad 
field only, and in this case the only possible nontrivial definition for the 
torsion tensor is given by $T^a\,_{\mu\nu}$. This torsion tensor is related to
the antisymmetric part of the Weitzenb\"ock  connection 
$\Gamma^\lambda_{\mu\nu}=e^{a\lambda}\partial_\mu e_{a\nu}$, which establishes
the Weitzenb\"ock spacetime. The curvature of the Weitzenb\"ock connection 
vanishes. However, the tetrad field also yields the metric tensor, which 
determines the Riemannian geometry. Therefore in the framework of a 
geometrical theory based on the tetrad field, one may use the concepts of 
both Riemannian and Weitzenb\"ock geometries. 

\section{TEGR with conformal invariance}

A conformal transformation on the space-time metric tensor $g_{\mu\nu}$ 
affects the length scales. It transforms $g_{\mu\nu}$ into 
$\tilde{g}_{\mu\nu}=e^{2\theta(x)}\,g_{\mu\nu}$, where $\theta(x)$ is an 
arbitrary function of the space-time coordinates. The same transformation on 
both the tetrad field and its inverse is defined by

\begin{equation}
\tilde{e}_{a\mu}=e^{\theta(x)}\,e_{a\mu}\,, \ \ \ \ \ \ \ \ \ \ \
\tilde{e}^{a\mu}=e^{-\theta(x)}\,e^{a\mu}\,.
\label{1}
\end{equation}
The expressions above determine the transformation properties of all 
geometrical quantities of interest in this analysis. The transformation of the 
projected components of the torsion tensor 
$T_{abc}=e_b\,^\mu e_c\,^\nu (\partial_\mu e_{a\nu}-\partial_\nu e_{a\mu})$ is
given by

\begin{eqnarray}
\tilde{T}_{abc}&=&
e^{-\theta}(T_{abc} + \eta_{ac}\,e_b\,^\mu \partial_\mu \theta
-\eta_{ab}\,e_c\,^\mu \partial_\mu \theta)\,, \nonumber \\
\tilde{T}^{abc}&=&
e^{-\theta}(T^{abc} + \eta^{ac}\,e^{b\mu} \partial_\mu \theta
-\eta^{ab}\,e^{c\mu} \partial_\mu \theta)\,. 
\label{2}
\end{eqnarray}
It follows from the equation above that for the trace of the torsion tensor
$T_a=T^b\,_{ba}$, we have

\begin{eqnarray}
\tilde{T}_a&=&e^{-\theta}(T_a-3\,e_a\,^\mu \partial_\mu \theta)\,, 
\nonumber \\
\tilde{T}^a&=&e^{-\theta}(T^a-3\,e^{a\mu} \partial_\mu \theta)\,.
\label{3}
\end{eqnarray}
With the help of equations (2) and (3), it is possible to verify that the 
behaviour of the three torsion squared terms that determine the Lagrangian 
density of teleparallel theories of gravity is given by

\begin{eqnarray}
\tilde{T}^{abc}\tilde{T}_{abc}&=& e^{-2\theta}(T^{abc}T_{abc}-
4T^\mu \partial_\mu\theta+6g^{\mu\nu}\partial_\mu\theta\partial_\nu \theta)\,,
\nonumber \\
\tilde{T}^{abc}\tilde{T}_{bac}&=& e^{-2\theta}(T^{abc}T_{bac}-
2T^\mu \partial_\mu\theta+3g^{\mu\nu}\partial_\mu\theta\partial_\nu \theta)\,,
\nonumber \\
\tilde{T}^a\tilde{T}_a&=&e^{-2\theta}(T^a\,T_a -6T^\mu \partial_\mu \theta
+9g^{\mu\nu}\partial_\mu\theta\partial_\nu \theta)\,.
\label{4}
\end{eqnarray}

In view of the equations above it is straightforward to check that the quantity

\begin{equation}
L={1\over 4}T^{abc}T_{abc}+{1\over 2}T^{abc}T_{bac}-{1\over 3}T^aT_a\,
\label{5}
\end{equation}
transforms under a conformal transformation according to 
$\tilde{L}=e^{-2\theta} L$. Equation (5) is the point of departure for the
construction of conformally invariant theories.

As a consequence of eq. (1) we find that for the determinant $e$ of the tetrad
field we have $\tilde{e} = e^{4\theta}\,e$. Therefore, we introduce a scalar 
field $\phi$ that is assumed to transform as

\begin{equation}
\tilde{\phi}=e^{-\theta} \phi\,.
\label{6}
\end{equation}
With the help of eq. (6) it is easy to verify that 

\begin{equation}
e\phi^2\biggl( 
{1\over 4}T^{abc}T_{abc}+{1\over 2}T^{abc}T_{bac}-{1\over 3}T^aT_a \biggr)
\label{7}
\end{equation}
is invariant under coordinate transformations and conformal transformations.
We note that in the TEGR \cite{Maluf1,Maluf2} the coefficient of the term 
$T^aT_a$ is $-1$, and not $-1/3$ as it appears in the expression above.

The Lagrangian density for the scalar field is established by noting that a 
covariant derivative may be defined as

\begin{equation}
D_\mu \phi=\biggl(\partial_\mu -{1\over 3} T_\mu \biggr)\phi\,,
\label{8}
\end{equation}
where $T_\mu=T^\lambda\,_{\lambda \mu}=T^a\,_{a\mu}$. Under the transformation
(1) we  have $\tilde{T}_\mu = e^{-\theta}(T_\mu-3\partial_\mu \theta)$,
and thus $\tilde{D}_\mu \tilde{\phi}=e^{-\theta}\,D_\mu \phi$.

Therefore the Lagrangian density

\begin{equation}
{\cal L} = k e\biggl[
-\phi^2\biggl( {1\over 4}T^{abc}T_{abc}+
{1\over 2}T^{abc}T_{bac}-{1\over 3}T^aT_a \biggr) 
+k'g^{\mu\nu}D_\mu \phi D_\nu \phi\biggr]\,,
\label{9}
\end{equation}
where $k=1/(16\pi G)$ is invariant under conformal transformations. The 
constant coefficient $k'$ is fixed by requiring that the condition $\phi = 1$
(or $\phi=\phi_0=$ constant) in ${\cal L}$ leads to the usual teleparallel 
Lagrangian that establishes the TEGR. It is easy to verify that for this
purpose we must have $k'=6$. 

Thus the Lagrangian density that is invariant under space-time
coordinate transformations and under conformal transformations, and that 
yields the Lagrangian density of the TEGR in the limiting case
$\phi \rightarrow \phi_0 =$ constant, is given by

\begin{equation}
{\cal L} = k e\biggl[
-\phi^2 \biggl( {1\over 4}T^{abc}T_{abc}+
{1\over 2}T^{abc}T_{bac}-{1\over 3}T^aT_a \biggr) 
+6g^{\mu\nu}D_\mu \phi D_\nu \phi\biggr]\,.
\label{10}
\end{equation}

The expression above may be simplified in two steps. First we rewrite the 
right hand side of (5) as 

\begin{equation}
{1\over 4}T^{abc}T_{abc}+{1\over 2}T^{abc}T_{bac}-{1\over 3}T^aT_a=
\Sigma^{abc}T_{abc} + {2\over 3}T^aT_a\,,
\label{11}
\end{equation}
where

\begin{equation}
\Sigma^{abc}={1\over 4} (T^{abc}+T^{bac}-T^{cab})
+{1\over 2}( \eta^{ac}T^b-\eta^{ab}T^c)\;,
\label{12}
\end{equation}
and 

\begin{equation}
\Sigma^{abc}T_{abc}=
{1\over 4}T^{abc}T_{abc}+{1\over 2}T^{abc}T_{bac}-T^aT_a
\label{13}
\end{equation}
is the scalar that establishes the Lagrangian of the TEGR 
\cite{Maluf1,Maluf2}, and ultimately yields Einstein's equations. Second, we
observe that

\begin{equation}
6g^{\mu\nu}D_\mu \phi D_\nu \phi=6g^{\mu\nu}\partial_\mu\phi\partial_\nu\phi
-4g^{\mu\nu}\phi (\partial_\mu \phi) T_\nu+
{2\over 3} \phi^2\,T^aT_a\,.
\label{14}
\end{equation}
By substituting eqs. (11) and (14) into eq. (10), we find

\begin{equation}
{\cal L}(e_{a\mu},\phi)=ke\biggl[-\phi^2\, \Sigma^{abc}T_{abc}+
6 g^{\mu\nu}(\partial_\mu \phi)(\partial_\nu \phi)
-4g^{\mu\nu}T_\nu \phi\,(\partial_\mu \phi)\biggr]\,.
\label{15}
\end{equation}
It is clear that by requiring $\tilde{\phi} =1$, for instance, by means of a
gauge conformal transformation, eq. (15) reduces to the Lagrangian density
of the TEGR. 

The Euler-Lagrange field equations obtained by varying 
${\cal L}(e_{a\mu},\phi)$ with respect
to $e^{a\mu}$ and $\phi$ are, respectively,

\begin{eqnarray}
&{}&e_{a\nu}e_{b\mu}\partial_\lambda(e\,\phi^2\, \Sigma^{b\nu \lambda })-
e\,\phi^2\,(\Sigma^{bc}\,_aT_{bc\mu}-{1\over 4}e_{a\mu}\Sigma^{bcd}T_{bcd})
\nonumber \\
&{}&-{3\over 2}e\,e_{a\mu}
g^{\lambda\nu}(\partial_\lambda \phi)(\partial_\nu\phi)
+3e\,e_a\,^\nu (\partial_\mu \phi)(\partial_\nu \phi)\nonumber \\
&{}&+e\,e_{a\mu}g^{\lambda \nu}T_\lambda \phi(\partial_\nu \phi)
-e\,\phi\, e_a\,^\nu(T_\nu \partial_\mu \phi + T_\mu \partial_\nu \phi)
\nonumber \\
&{}&-e\,g^{\lambda \nu}\,\phi(\partial_\lambda \phi) T_{a\mu\nu} \nonumber \\
&{}&-e_{a\nu}e_{b\mu} \partial_\sigma
\lbrack e\,g^{\lambda \nu} \phi(\partial_\lambda \phi)e^{b\sigma}\rbrack+
e_{a\sigma}e_{b\mu} \partial_\nu
\lbrack e\,g^{\lambda \nu} \phi(\partial_\lambda \phi)e^{b\sigma}\rbrack=0\,,
\label{16}
\end{eqnarray}
and

\begin{equation}
\partial_\nu(e\,g^{\mu\nu}\partial_\mu \phi)+{1\over 6}
\phi \lbrack e\, \Sigma^{abc}T_{abc}-2\partial_\mu (e\,T^\mu)\rbrack=0\,.
\label{17}
\end{equation}
The field equation for $\phi$ is obtained exactly in the form presented 
above. However, we note that the last two terms of the equation (the 
terms between brackets) define a quantity that is {\it identically} 
equal to the scalar curvature constructed out of the tetrad field only
\cite{Maluf2},

\begin{equation}
e\, \Sigma^{abc}T_{abc}-2\partial_\mu (e\,T^\mu)\equiv -e\,R(e)\,.
\label{18}
\end{equation}
It is precisely this identity that establishes the equivalence between the
field equations of the TEGR and Einstein's equations.
Therefore the final form of the field equation for $\phi$ is

\begin{equation}
\partial_\nu(e\,g^{\mu\nu}\partial_\mu \phi)-{1\over 6}e\,\phi R(e)=0\,.
\label{19}
\end{equation}

The solutions of the field equations (16) and (19) are related to the 
solutions of Einstein's equations in vacuum. In order to prove this
statement, we need first to verify whether $\phi = \phi_0 =$ constant is a 
solution of the field equation (19). 

By making $\phi=1$, for instance, in eq. (16), the latter is reduced to

\begin{equation}
e_{a\nu}e_{b\mu}\partial_\lambda(e\,\Sigma^{b\nu \lambda })-
e\,(\Sigma^{bc}\,_aT_{bc\mu}-{1\over 4}e_{a\mu}\Sigma^{bcd}T_{bcd})=0\,.
\label{20}
\end{equation}
This equation is equivalent to Einstein's equations in vacuum. The left hand 
side of the equation is identically equal to 
$1/2\,\lbrack(e R_{a\mu}-{1\over 2} e_{a\mu}R) \rbrack$ \cite{Maluf2}.
A solution of the field equation (19) of the type $\phi=\phi_0=$ constant 
is possible provided $R=0$. However, it follows
from the Einstein's equations in vacuum that the scalar curvature $R$ vanishes.
Therefore the set $(\phi_0, e_{a\mu})$, where $\phi_0=$ constant and $e_{a\mu}$
is solution of Einstein's equations in vacuum, is solution of the field 
equations (16) and (19). 

Since $\phi=1$ is a possible solution of the field equations, integration by
parts in the Lagrangian density (15) is not a trivial and straightforward 
procedure. By performing integration by parts in the last term of (15), 
namely, in $-4e\,g^{\mu\nu}T_\nu \phi\,(\partial_\mu \phi)$, the surface term 
that arises in the procedure does not vanish in the context of asymptotically
flat space-times. Yet, neglecting the nonvanishing of this term, 
integration by parts in the term above yields 
$2\partial_\mu (e\,T^\mu)\,\phi^2$. By adding to the latter the first term of 
(15), i.e., $-e\phi^2\,\Sigma^{abc}T_{abc}$, and making use of the identity
(18), we conclude that the Lagrangian density (15) is precisely given by
$e\lbrack\phi^2\,R+6 g^{\mu\nu}(\partial_\mu\phi)(\partial_\nu \phi)\rbrack$.
Therefore the theory defined by (10) or (15) is equivalent to the usual
formulation of the Hilbert-Einstein Lagrangian endowed with conformal symmetry,
but {\it not} the theory defined by (9) with $k' \ne 6$. Therefore theories
for which $k'\ne 6$ in eq. (9) are conformal theories of gravity that have
not been considered so far.

\section{A purely geometrical conformal theory}

A more complicated geometrical construction of a theory of gravity with
conformal invariance is based on the square of the quantity given by eq. (5).
The theory defined by

\begin{equation}
{\cal L}(e_{a\mu})=\alpha e  L^2\,,
\label{21}
\end{equation}
where $\alpha$ is a dimensionless constant, is invariant under conformal
transformations. No scalar field is needed for the conformal symmetry. 
Moreover, it is not possible to envisage any possible relation between $L^2$
and $C^{\alpha\beta\mu\nu}C_{\alpha\beta\mu\nu}$, where 
$C_{\alpha\beta\mu\nu}$ is the Weyl tensor, since it is not possible to
write $L$ given by (5) in terms of the metric tensor. The field equation 
obtained by varying eq. (21) is given by

\begin{equation}
e_{a\nu}e_{b\mu}\partial_\lambda(e\,L\,\Lambda^{b\nu\lambda})-
e\,L(\Lambda^{bc}\,_a T_{bc\mu}-{1\over 8}e_{a\mu}\,L)=0\,,
\label{22}
\end{equation}
where

\begin{eqnarray}
\Lambda^{abc}&=&\Sigma^{abc}+{1\over 3}(\eta^{ab}T^c-\eta^{ac}T^b)
\nonumber \\
&=& {1\over4}(T^{abc}+T^{bac}-T^{cab})+
{1\over 6}(\eta^{ac}T^b-\eta^{ab}T^c)\,.
\label{23}
\end{eqnarray}
By contracting eq. (22) with $e^{a\mu}$, it is not difficult to see that the
equation is trace free, as it is expected to be.
Evidently the field equation (22) is quite intricate, and it is not easy
to find an analytic solution of the equation. 

The Lagrangian density given by eq. (21) is not the only combination of the
torsion tensor that yields a conformally invariant action integral. By 
inspecting eq. (4) it is easy to verify that 

\begin{equation}
L_1=A\, T^{abc}T_{abc}+ B\, T^{abc}T_{bac}+ C\, T^aT_a\,,
\label{24}
\end{equation}
where $A$, $B$, and $C$ are constants that satisfy

\begin{equation}
2A+B+3C=0\,,
\label{25}
\end{equation}
transforms covariantly under conformal transformations:
$\tilde{L}_1=e^{2\theta} L_1$. Therefore the family of theories defined by the
Lagrangian density

\begin{equation}
{\cal L}(e_{a\mu})= e  L_1 L_2 \,,
\label{26}
\end{equation}
where $L_2=A'\, T^{abc}T_{abc}+ B'\,T^{abc}T_{bac}+ C'\, T^aT_a$, and $A'$, $B'$ 
and are $C'$ constants that also satisfy $2A' +B' +3C'=0$, is invariant under 
conformal transformations.

\section{ Final remarks}

In this article we have presented families of conformal theories of gravity
constructed out of the torsion tensor. The theory defined by the Lagrangian
densities (10) or (15) is equivalent to the Hilbert-Einstein action endowed
with conformal symmetry. The theory defined by eq. (9), with $k' \ne 6$,
deviates from the standard formulation of general relativity. In the 
Lagrangians (9), (10) and (15) a scalar field is necessary in order to
ensure the conformal symmetry.

On the other hand, the theory defined by eq. (21) is genuinely geometrical. 
Moreover it is not equivalent to the usual Weyl Lagrangian density. There
is no geometrical relation between $L^2$, which is a functional of 
$e_{a\mu}$, and the square of the Weyl tensor $C_{\alpha\beta\mu\nu}$, 
constructed out of the metric tensor.
By combining eqs. (11) and (18) we see that

\begin{equation}
L=-R+{2\over e}\partial_\mu (eT^\mu)+{2\over 3}T_\mu T^\mu\,,
\label{27}
\end{equation}
where $R$ is the scalar curvature.
It is impossible to write a vector quantity like $T^\mu$ as a covariant
functional constructed out of the second rank tensor $g_{\mu\nu}$. 

For the sake of completeness, we will indicate the construction of the 
teleparallel version of the Weyl Lagrangian density. Let us consider the
Weitzenb\"{o}ck connection 
$\Gamma^\lambda_{\mu\nu}=e^{a\lambda}\partial_\mu e_{a\nu}$, the 
Christofell symbols $^0\Gamma^\lambda_{\mu\nu}$ and the contorsion tensor
$K_\mu\,^\lambda\,_\nu$ defined by

$$K_\mu\,^\lambda\,_\nu={1\over 2}(T^\lambda\,_{\mu\nu}+T_\mu\,^\lambda\,_\nu
+T_\nu\,^\lambda\,_\mu)\,,$$
where $T^\lambda\,_{\mu\nu}=e^{a\lambda}(\partial_\mu e_{a\nu}
-\partial_\nu e_{a\mu})$. These quantities are {\it identically} related by

\begin{equation}
\Gamma^\lambda_{\mu\nu}=\;^0\Gamma^\lambda_{\mu\nu} + K_\mu\,^\lambda\,_\nu\,.
\label{28}
\end{equation}
The curvature tensor constructed out of the Weitzenb\"{o}ck connection, on the
left hand side of eq. (28),
vanishes identically. Thus by substituting this connection into the standard 
expression of the Riemann tensor,

$$R^\lambda\,_{\sigma\mu\nu}=\partial_\mu \Gamma^\lambda_{\sigma\nu}-
\partial_\nu\Gamma^\lambda_{\sigma\mu}
+\Gamma^\lambda_{\beta\mu}\Gamma^\beta_{\sigma\nu}
-\Gamma^\lambda_{\beta\nu}\Gamma^\beta_{\sigma\mu}\,,$$
we arrive at the identity

\begin{equation}
R^\lambda\,_{\sigma\mu\nu}(^0\Gamma)=-\nabla_\mu K_\sigma\,^\lambda\,_\nu+
\nabla_\nu K_\sigma\,^\lambda\,_\mu - 
K_\beta\,^\lambda\,_\mu K_\sigma\,^\beta\,_\nu+
K_\beta\,^\lambda\,_\nu K_\sigma\,^\beta\,_\mu\,,
\label{29}
\end{equation}
where the left hand side is the Riemann-Christoffel tensor. The covariant
derivative $\nabla_\mu$ is constructed out of the Christoffel symbols 
$^0\Gamma^\lambda_{\mu\nu}$. It is clear that substitution of the left
hand side of the expression above into the Weyl Lagrangian density,

\begin{equation}
\sqrt{-g}\,C^{\alpha\beta\mu\nu}C_{\alpha\beta\mu\nu}=\sqrt{-g} 
\,(R^{\alpha\beta\mu\nu}R_{\alpha\beta\mu\nu}-2R^{\mu\nu}R_{\mu\nu}+
{1\over 3}R^2)\,,
\label{30}
\end{equation}
or even into the simplified, reduced form 

\begin{equation}
\sqrt{-g}\,C^{\alpha\beta\mu\nu}C_{\alpha\beta\mu\nu}=-2\sqrt{-g}
(R^{\mu\nu}R_{\mu\nu}-{1\over 3}R^2)\,,
\label{31}
\end{equation}
obtained with the help of the Lanczos identity, yields a very complicated 
structure  of a conformally invariant teleparallel theory, which contains 
derivatives of the torsion tensor, and is in no way related to expression 
(21) or to the square of eq. (27).

Therefore the Weyl Lagrangian density is not the 
unique purely geometrical entity that yields a conformal theory of gravity.
The Lagrangian density given by eq. (26) seems to be the most general 
family of theories constructed out of the torsion tensor $T_{a\mu\nu}$ only
(i.e., with no derivatives of $T_{a\mu\nu}$)
that are invariant under conformal transformations. The theories established
here may play a role in the formulation of the quantum theory of gravity.\par
\bigskip
\noindent {\bf Acknowledgements}\par
\bigskip
\noindent The authors are grateful to Dr. N. Tamanini for pointing out 
the feature displayed by eqs. (24) and (25).

\end{document}